\documentclass[12pt]{iopart}
\usepackage{nccfoots}
\usepackage{graphicx}
\usepackage{braket}
\usepackage{color}
\newcommand{\opdu}[3]{|#1\rangle_#2\langle#3|}

\definecolor{mygreen}{rgb}{0,0.5,0}
\definecolor{myblue}{rgb}{0,0,0.75}
\definecolor{mymagenta}{cmyk}{0,1,0,0.12}
\usepackage[colorlinks=true,
    linkcolor=blue,
    citecolor=blue,
    urlcolor =blue]{hyperref}

\usepackage{iopams}  
\expandafter\let\csname equation*\endcsname\relax

\expandafter\let\csname endequation*\endcsname\relax

\usepackage{amsmath}
\begin{document}
\title{Robust quantum state transfer via topologically protected edge channels in dipolar arrays}

\renewcommand\footnotemark{}
\renewcommand\footnoterule{}
\author{C. Dlaska$^\dagger$\footnotetext{These two authors contributed equally}, B. Vermersch$^\dagger$ and P. Zoller}

\address{Institute for Quantum Optics and Quantum Information of the Austrian
Academy of Sciences, A-6020 Innsbruck, Austria}
\address{Institute for Theoretical Physics, University of Innsbruck, A-6020
Innsbruck, Austria}
\vspace{10pt}

\begin{abstract}
We show how to realize quantum state transfer between distant qubits using the chiral edge states of a two-dimensional topological spin system. Our implementation based on Rydberg atoms allows to realize the quantum state transfer protocol in state of the art experimental setups. In particular, we show how to adapt the standard state transfer protocol to make it robust against dispersive and disorder effects.
\end{abstract}
\section{Introduction}

Quantum state transfer aims at transferring the state of a first to a second qubit, i.e.~$\left(A \ket{0}_1   + B \ket{1}_1 \right)   \otimes \ket{0}_2 \to \ket{0}_1 \otimes  \left(A \ket{0}_2  + B \ket{1}_2\right)$, and represents a basic building block of quantum communication and quantum information processing in a quantum network~\cite{Cirac1997,Kimble2008}. Such a quantum network consists of nodes representing qubits as quantum memory, or in a broader context quantum computers, which are connected by quantum channels. Quantum networks are discussed both as local quantum networks connecting, and thus scaling up small scale quantum computers~\cite{Hucul2014,Northup:2014gv,Nickerson2014}, or in quantum communication between distant nodes~\cite{Cirac1997,Reiserer2015}.

\begin{figure}[h!]
\centering
\includegraphics[width=\textwidth]{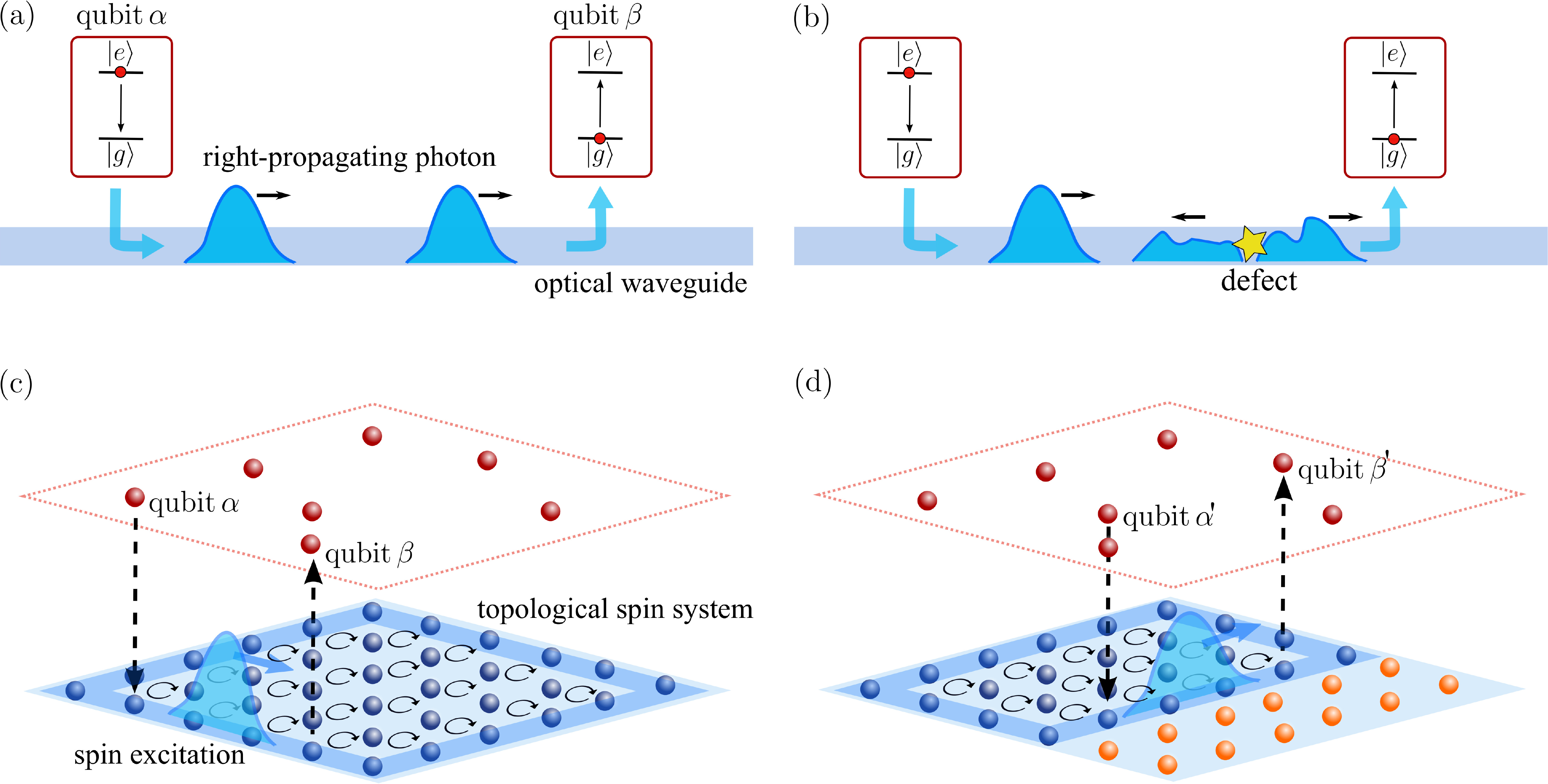}
\caption{Quantum state transfer between two distant qubits. (a) Using a bidirectional waveguide as a quantum channel the process is achieved by chiral emission of a right-propagating photon from the qubit $\alpha$, which is then absorbed by the second qubit $\beta$. 
(b) The presence of defects in the waveguide can distort the photon shape and induce back-scattering, affecting the fidelity of the quantum state transfer.
(c) The chiral edge states of a topological spin system are protected against back-scattering. In this case, the chiral edge states act as a unidirectional quantum channel and quantum state transfer between the two qubits is mediated by a spin excitation propagating along the edge of the material.
(d) In our Rydberg dressing implementation, interactions between the atoms of the topological spin system can be switched on and off dynamically via time-dependent addressing laser beams, allowing to reshape the edges of the material ``on-demand'' to connect arbitrary qubits.\label{fig:intro}}
\end{figure}

The goal in a physical implementation of quantum state transfer is to achieve transmission of a qubit with high fidelity through the quantum channel, i.e.~avoiding decoherence. In a wide area quantum network the natural carrier of quantum information  will be photons as ``flying qubits'' propagating in fibers or in free space, as a physical realization of the quantum channel, where a quantum optical interface allows storage in ``stationary qubits'' represented by two-level atoms as quantum memory~\cite{Goban2014,Tiecke2014}. Quantum state transfer between atoms stored in high-Q cavities connected by a photonic channel was reported in seminal experiments~\cite{Ritter2012,Nolleke2013} following the protocol described in~\cite{Cirac1997}. A remarkable recent experimental development has been {\em chiral quantum interfaces}  \cite{Mitsch2014,Sollner2015,Young2015,Coles2016} in the context of chiral quantum optics~\cite{Lodahl2016}, where two-level systems coupled to 1D photonic nanostructures or nanofibers, control the direction of propagation of emitted photons with a {\em chiral light-matter coupling}. This is illustrated in Figure~\ref{fig:intro}(a) as a two-level atom representing a first qubit ($\alpha=1$) $\ket{0}\equiv\ket{g}$, $\ket{1}\equiv\ket{e}$, which decays from the excited state $\ket{e}$ to the ground state $\ket{g}$, emitting a photon into a 1D waveguide traveling unidirectionally to the right, which then can be restored into the second atom or qubit ($\beta=2$) achieving quantum state transfer. In nanophotonics this chiral coupling occurs naturally due to  spin-orbit coupling of light~\cite{Lodahl2016}. In local area or ``on-chip'' chiral quantum networks, quantum state transfer can also be achieved via 1D phonon and spin channels~\cite{Ramos2016}. In the latter case, magnons as spin excitations take the role of the ``flying qubit''~\cite{Bose2003}, and a physical implementation of a chiral quantum interfaces between spins has been recently given in~\cite{Vermersch2016} for setups of Rydberg atoms arranged as 1D strings and ion chains. 

In chiral quantum optics the chiral light-matter interface selects the propagation direction of the traveling photon (or spin) wavepacket [c.f.~Figure~\ref{fig:intro}(a)], while the 1D waveguide supports both right and left going modes. Such a setup will thus not be protected against back-scattering from imperfections in the waveguide, as illustrated in ~Figure~\ref{fig:intro}(b). Instead we will be interested below in chiral quantum channels arising as chiral edge channels in 2D topological quantum materials~\cite{Yao2013,Yang2016} (c.f.~Figure~\ref{fig:intro}(c)). Such topological quantum materials can be realized in condensed matter~\cite{Hasan2010,Qi2011}, and be engineered as synthetic quantum matter with atomic systems~\cite{Aidelsburger2013,Miyake2013,Mancini2015,Stuhl2015,Goldman2016}, or in photonic setups~\cite{Hafezi2013,Rechtsman2013}. The distinguishing property of chiral edge channels is that they allow only unidirectional propagation around the edge of the topological quantum material. In the context of quantum state transfer coupling a qubit to a chiral edge channel will thus not only provide us with an {\em a priori} chiral qubit-channel interface, but chiral edge channels are by their very nature immune against back-scattering from defects~\cite{Hasan2010,Qi2011}. 

It is the purpose of the present work to study quantum state transfer of qubits via chiral quantum edge channels in a physical setting provided by dipolar arrays of Rydberg atoms. Motivated by recent experiments demonstrating dipolar Rydberg~\cite{Barredo2015,Maller2015,Weber2015,Zeiher2016,Browaeys2016} and polar molecules~\cite{Yan2013} arrays, and building on recent proposals to engineer topological phases in spin systems realized by Rydberg atoms or polar molecules~\cite{Yao2012,Syzranov2014,Peter2015}, 
we propose an implementation where the dipolar interactions are realized by weakly admixing ground-state atoms to Rydberg states via laser fields~\cite{Santos2000,Pupillo2010,Henkel2010}.
A key property of such a Rydberg dressing implementation is that - with an appropriate spatial laser addressing - we can rearrange the edge of our engineered topological material, i.e.~we can dynamically shape the edge channels to connect arbitrary pairs of qubits [c.f.~Figure~\ref{fig:intro}(d)].
Furthermore, our implementation provides a framework for illustrating in an experimentally realistic setup various features of quantum state transfer via topologically protected, and thus robust quantum channels, but also for realizing a spectroscopy of chiral edge channels {\em per se}. Thus we show how the measurement,  by the qubits, of edge state wave-packets allows to realize high fidelity quantum state transfers, robust against disorder and dispersive effects~\cite{Ramos2016,nikolopoulos2013quantum}.

Our manuscript is organized as follows.
First, in Section~\ref{sec:model}, we introduce our model of a quantum network with qubits connected via topological quantum channels.
In Section~\ref{sec:implementation}, we present an implementation of this model based on Rydberg-dressed atoms.
We then numerically study in Section~\ref{sec:statetransfer} the robustness of the quantum state transfer protocol and the role of disorder and dispersive effects.
Finally, in Section~\ref{sec:chirp}, we propose and assess the performance of a protocol, which exploits the chiral properties of the edge states, to achieve a perfect quantum state transfer, robust against static imperfections.

\section{Model of a topological network and quantum state transfer using chiral edge states}\label{sec:model}

\begin{center}
\begin{figure}[h!]
\includegraphics[width=\textwidth]{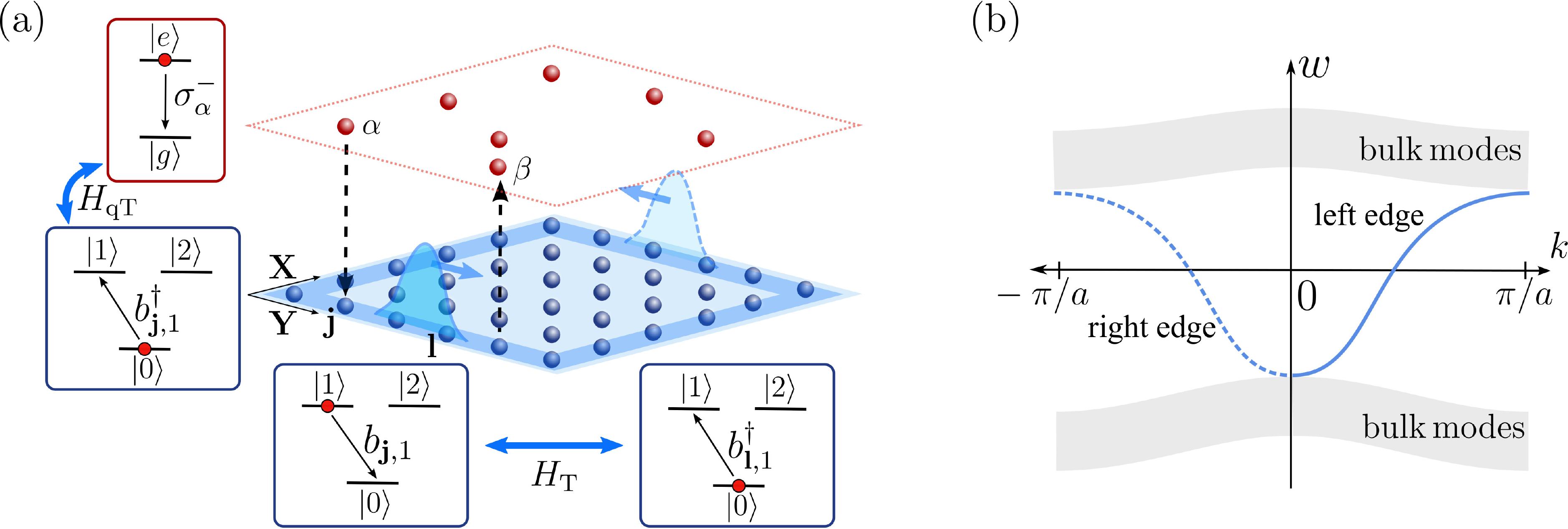}
\caption{Model of qubits coupled to a topological spin system. (a) The qubits are coupled to three-level systems. (b) Typical dispersion relation $\omega_m(k)$ of a topological spin system. The edge mode (blue line) propagates with positive (negative) velocity along the left (right) edge.\label{fig:concept}}
\end{figure}
\end{center}

In this section we present our model of qubits, represented by a set of two-level atoms or spin-$\tfrac{1}{2}$, which can be coupled to an engineered 2D topological spin system~\cite{Hasan2010,Qi2011}. The setup we have in mind is illustrated in Figure~\ref{fig:concept}(a): the qubits are arranged on a quantum memory layer above the topological spin system (TSS), where chiral edge states play the role of the quantum channel.
While in the present section we will define this model on a abstract level, we will discuss a physical implementation of both qubits and the topological spin system with Rydberg atoms in Section~\ref{sec:implementation}.

Quantum state transfer between a chosen pair of qubits $\alpha=1$ and $\beta=2$ on the quantum memory layer can be achieved by first arranging the dynamical chiral edge channels to connect the qubit pair, and - again using lasers and dipolar interactions - swapping the state of the first qubit to a wave packet propagating in the edge channel, where it can be restored in the second qubit. This is illustrated in Figure~\ref{fig:intro}(a) for a pair of qubits at the edge of a square topological spin system. 

We emphasize that we consider in this work the limit of zero temperature where the precise control of the state of the quantum channel, which is not affected by the presence of thermal excitations, allows to achieve the quantum state transfer.
This assumption is valid in particular for the Rydberg-dressing implementation presented in the next section, where the spin state of the TSS atoms are controlled with excellent precision~\cite{Saffman2010,Browaeys2016}. In contrast, for other platforms using for instance microwave  waveguides or mechanical resonator arrays as quantum channel, temperature effects have to be included for a realistic description of the quantum state transfer~\cite{Habraken2012}.

The Hamiltonian associated with our model consists of three parts
\begin{equation}
H=H_\mathrm{q}+H_\mathrm{T}+H_\mathrm{qT}\label{eq:H},
\end{equation}
corresponding respectively to the qubits, the topological spin system and the coupling between them. 
In the following, we first present the Hamiltonian of the qubits and the topological spin system (Section~\ref{sec:Ham}) and then we show how to describe the coupling of the qubits to the chiral edge states of the topological spin system  (Section~\ref{sec:coupling}).
Finally, we present the quantum state transfer protocol in our setting (Section~\ref{sec:protocol}).

\subsection{Two-level systems and topological spin system Hamiltonian}\label{sec:Ham}

The qubits $\alpha=1,..,N_\mathrm{q}$ forming the quantum memory layer, with ground state $\ket{g}$ and excited state $\ket{e}$, are located at positions $(x_\alpha,y_\alpha,z_\alpha) $. The qubit Hamiltonian is given by
\begin{equation}
H_\mathrm{q} = \Delta_\mathrm{q}\sum_{\alpha=1}^{N_\mathrm{q}}\sigma_\alpha^{+}\sigma_\alpha^{-},\label{eq:Hqubit}
\end{equation}
with $\sigma_\alpha^{-}=|g\rangle_\alpha\langle e|$, $\sigma_\alpha^{+}=\left(\sigma_\alpha^{-}\right)^\dagger$ and energy offsets $\Delta_\mathrm{q}$.

 The topological spin system is represented by V-type  three-level systems $\mathbf{j}=(j_x,j_y)$ with $j_x=1,..,N_\mathrm{X},\ j_y=1,..,N_\mathrm{Y}$, where $\ket{0}_\mathbf{j}$ denotes the ground state and $\ket{1}_\mathbf{j}\ket{2}_\mathbf{j}$ the two excitation states of atom $\mathbf{j}$, which are placed at fixed positions $(x_\mathbf{j},y_\mathbf{j},z_\mathbf{j}=0)$ in the $X,Y$ plane according to a square lattice of lattice spacing $a$. Similar to the Harper Hofstadter Hamiltonian~\cite{Harper1955,Hofstadter1976}, which has been recently realized in cold atom experiments~\cite{Aidelsburger2013,Miyake2013,Mancini2015,Stuhl2015}, a topological band structure is obtained by allowing the spin excitations  ($\ket{1}$,$\ket{2}$), aka magnons, to acquire a phase when hopping around a closed loop in the lattice. The corresponding Hamiltonian is given by
\begin{equation}
H_\mathrm{T} =\sum_{\mathbf{j}}\sum_{\nu=1,2}\delta_\nu b_{\mathbf{j},\nu}^\dagger b_{\mathbf{j},\nu} + \sum_{\mathbf{j}\neq \mathbf{l}}\mathbf{b_\mathbf{j}}^\dagger  h(\mathbf{r}_{\mathbf{j},\mathbf{l}}) \mathbf{b_\mathrm{l}}. 
\label{eq:HBreal}
\end{equation}
Here, we map the spin excitations to hard-core boson particles where the two-element vector $\mathbf{b_\mathrm{\mathbf{j}}}=[b_{\mathbf{j},1},b_{\mathbf{j},2}]$, with $b_{\mathbf{j},\nu}=|0\rangle_\mathbf{j}\langle\nu|$, represents the two (hard-core boson) annihilation operators of the excitations at site $\mathbf{j}$ and $\delta_\nu$ is the corresponding energy offset. 
The matrix  $h(\mathbf{r_{j,l}})$, with  $\mathbf{r_{j,l}}=(x_\mathbf{j}-x_\mathbf{l})\mathbf{X}+(y_\mathbf{j}-y_\mathbf{l})\mathbf{Y}$, describes the transfer of an excitation between sites $\mathbf{j}$ and $\mathbf{l}$ and can be written as 
\begin{equation}
h(\mathbf{r_{j,l}})=
\left[ \begin{array}{cccc}
t_1(\mathbf{r_{j,l}}) & w(\mathbf{r_{j,l}})e^{-i\phi(\mathbf{r_{j,l})}}  \\
w(\mathbf{r_{j,l}})e^{i\phi(\mathbf{r_{j,l}})}   & t_2(\mathbf{r_{j,l}})
 \end{array} \right],
 \label{eq:h}
\end{equation}
where the phases $\phi(\mathbf{r_{j,l}})$ are responsible for the existence of Quantum Hall topological bands characterized by non-vanishing Chern numbers~\cite{Syzranov2014,Peter2015} and thus chiral edge states~\cite{Hasan2010,Qi2011,Peter2015} [c.f.~Figure~\ref{fig:concept}(a)].

By considering for convenience the limit of an infinite number of atoms in one direction, $N_\mathrm{Y}\to\infty$,  we obtain the dispersion relation of bulk and edge modes
 by Fourier transforming along the $Y$ axis and diagonalizing the TSS Hamiltonian [c.f.~\ref{app:kycode}],\begin{equation}
H_\mathrm{T} = \sum_{m}\int_{-\pi/a}^{\pi/a}\mathrm{d}k\,\omega_m(k) b_{k,m}^\dagger b_{k,m},
\label{eq:HBk1}
\end{equation}
where $\omega_m(k)$ denotes the dispersion relation corresponding to the TSS eigenmode $b_{k,m}$~\footnote{This dispersion relation is only valid in the case of a single TSS excitation, where we can neglect the hard-core character of the boson operators $b_\mathbf{j}$.}. Figure~\ref{fig:concept}(b) shows a typical example of dispersion relation curves $\omega_m(k)$.
The edge state dispersion relation corresponding to $m=\mathrm{a}$ $\omega_\mathrm{a}(k)$ (blue line) represents localized modes propagating with group velocity $v_\mathrm{a}(k)\equiv \partial\omega_\mathrm{a}(k)/\partial k$.
Their chirality originates from the fact that the modes propagating with positive velocity are located on the left edge, while the modes moving in the other direction are located on the right edge [c.f.~Figure~\ref{fig:concept}(a)]. At this point we want to emphasize that the absence of counter-propagating modes (for example with a negative velocity on the left edge) guarantees the absence of reflections originated from local defects. We analyze in more details the robustness of the edge state channels in Sections~\ref{sec:statetransfer} and \ref{sec:chirp}.

\subsection{Coupling between qubits and topological spin system}\label{sec:coupling}
The last part of the Hamiltonian $H$ of our model corresponds to the coupling between the qubits and the topological spin system. To achieve the quantum state transfer, we are interested in coupling the qubits predominantly to the edge modes. This is achieved by positioning the qubits in the vicinity of the edge of the topological spin system and choosing the qubit transition frequency $\Delta_\mathrm{q}$ to match with the energy of an edge state $b_{\bar k_\mathrm{a},\mathrm{a}}$, where the resonant wave-vector $\bar k_\mathrm{a}$ is defined via $\omega_\mathrm{a}(\bar k_\mathrm{a})=\Delta_\mathrm{q}$. 
The coupling Hamiltonian is written as
\begin{equation}
H_\mathrm{qT} = \sum_{\mathbf{j},\alpha,\nu}g_{\nu}(\mathbf{r}_{\mathbf{j},\alpha},t) b_{\mathbf{j},\nu}^\dagger\sigma_\alpha^- +\mathrm{h.c.},
\label{eq:HqB_real}
\end{equation}
where $g_{\nu}(\mathbf{r_{j,\alpha}},t)$ represents the coupling of an excitation $\ket{e}_\alpha$ of the qubit $\alpha$ to a TSS excitation at site $\mathbf{j}$, encoded in one of the two levels $\nu=1,2$ and depends on the relative vector $\mathbf{r}_{\mathbf{j},\alpha}=(x_\mathbf{j}-x_\alpha)\mathbf{X}+(y_\mathbf{j}-y_\alpha)\mathbf{Y}-z_\alpha\mathbf{Z}$. Moreover, we consider the coupling terms $g_{\nu}(\mathbf{r_{j,\alpha}},t)$ to be time-dependent, which allows to form wave-packets of edge modes propagating with a well-defined shape~\cite{Cirac1997,Gorshkov2007b}. In the basis of the eigenmodes $b_{k,m}$, the coupling Hamiltonian takes the form:
\begin{equation}
H_\mathrm{qT} = \sum_{m,\alpha}\int\mathrm{d}k\, g_{k,m}^{(\alpha)}(t) e^{-ik y_\alpha}b_{k,m}^\dagger\sigma_\alpha^- +\mathrm{h.c.},
\end{equation}
with the coupling strength $g_{k,m}^{(\alpha)}(t)=(1/\sqrt{2\pi})\sum_{\mathbf{j},\nu}e^{-ik (y_\mathbf{j}-y_\alpha)}g_\nu (\mathbf{r_{j,\alpha}},t) c_{x_\mathbf{j},\nu}^{(k,m)}$, where the coefficients $c_{x_\mathbf{j},\nu}^{(k,m)}$ describe the spatial properties of the eigenmodes and are given in \ref{app:kycode}.

\subsection{Quantum state transfer}\label{sec:protocol}

Let us now apply our model to realize a quantum state transfer protocol~\cite{Cirac1997} using chiral edge states~\cite{Yao2013}. The formal process we want to achieve is the transfer of any superposition state $\ket{s}=A\ket{g}+B\ket{e}$ of a qubit $\alpha=1$ to a second qubit $\beta=2$ mediated by a wave-packet propagating in the quantum channel [c.f.~Figure~\ref{fig:intro}(a),(c)]:
 \begin{eqnarray}
\left[A\ket{g}_1+B\ket{e}_1\right]  \ket{0}_\mathrm{T}   \ket{g}_2
\nonumber \\
 \to   \ket{g}_1 \left[ A \ket{0}_\mathrm{T}    + B  \int \mathrm{d}k\,c_{k,\mathrm{a}} (t) e^{-i\omega_m(k)t} b^\dagger _{k,\mathrm{a}} \ket{0}_\mathrm{T} \right]  \ket{g}_2
 \nonumber \\
 \to \ket{g}_1  \ket{0}_\mathrm{T}  \left[A\ket{g}_2+B\ket{e}_2\right]   , 
 \end{eqnarray}
with $\ket{0}_\mathrm{T}=\prod_{j}\ket{0}_\mathbf{j}$ the excitation vacuum of the topological spin system.
In order to derive the form of the coupling $g_{k,m}^{(\alpha)}(t)$, which achieves the quantum state transfer, we write the general wave-function
\begin{equation}
\ket{\psi(t)}=\left(\sum_\alpha c_{e,\alpha}(t)e^{-i\Delta_\mathrm{q}t}\sigma_\alpha^+ +\sum_m\int\mathrm{d}k\,c_{k,m}(t)e^{-i\omega_m(k)t}b_{k,m}^\dagger\right)\ket{V},\label{eq:WW}
\end{equation}
describing the propagation of a single excitation in the total system, with $\ket{V}\equiv\ket{0}_\mathrm{T}\otimes\prod_\alpha\ket{g}_\alpha$. As presented in \ref{app:WW}, the Wigner-Weisskopf treatment, valid in the weak-coupling regime $h(\mathbf{r_{j,l}})\gg g_\nu(\mathbf{r}_{\mathbf{j},\alpha})$, eliminates the contribution of the TSS, resulting in the following set of equations for the qubit amplitudes
\begin{eqnarray}
\dot{c}_{e,1}(t)&=&-\frac{1}{2}\gamma_{\mathrm{a},1}(t) c_{e,1}(t)\label{eq:gamma1}\\
\dot{c}_{e,2}(t)&=&-\frac{1}{2}\gamma_{\mathrm{a},2}(t) c_{e,2}(t)\nonumber\\
&&-\sqrt{\gamma_{\mathrm{a},1}(t-\tau_\mathrm{a})\gamma_{\mathrm{a},2}(t)} e^{i(\bar{k}_\mathrm{a}d-(\eta_2-\eta_1))}c_{e,1}(t-\tau_\mathrm{a}),\label{eq:gamma2}
\end{eqnarray}
where 
\begin{equation}
\gamma_{\mathrm{a},\alpha}(t)=2\pi\frac{|g_{\bar{k}_\mathrm{a},\mathrm{a}}^{(\alpha)}(t)|^2}{|v_\mathrm{a}(\bar{k}_\mathrm{a})|}\label{eq:gammaa}
\end{equation}
denotes the coupling of the qubit $\alpha$ to the edge mode $m=\mathrm{a}$, $g_{k,\mathrm{a}}^{(\alpha)}(t)=|g_{k,\mathrm{a}}^{(\alpha)}(t)|e^{i\eta_\alpha}$, $d=y_2-y_1$ is the distance between the two qubits along the $Y$ axis and $\tau_\mathrm{a}=d/|v_\mathrm{a}(\bar{k}_\mathrm{a})|$ represents the time delay between the qubits.
We emphasize that equations \eref{eq:gamma1} and \eref{eq:gamma2} illustrate the unidirectional coupling between the two qubits.
Similar to the original proposal~\cite{Cirac1997}, the quantum state transfer protocol can be then realized with the time-dependent coupling~\cite{Gorshkov2007b,Stannigel2011}
 \begin{equation}
\gamma_{\mathrm{a},1}(t)=\frac{2\sqrt{c}\gamma_{0}e^{-c(t-t_{0})^{2}}}{2\sqrt{c}-\sqrt{\pi}\gamma_{0}\mathrm{erf}\left(\sqrt{c}(t-t_{0})\right)},\label{eq:pulse}
 \end{equation}
which generates a Gaussian edge state wave-packet of temporal width $\sqrt{c}$. This symmetric pulse can be then reabsorbed by the second qubit  via a time-reversed pulse $\gamma_{\mathrm{a},2}(t)=\gamma_{\mathrm{a},1}(\tau_\mathrm{a}-t)$. 

\section{Implementation with Rydberg atoms}\label{sec:implementation}
\begin{figure}[h]
\centering
\includegraphics[width=0.99\columnwidth]{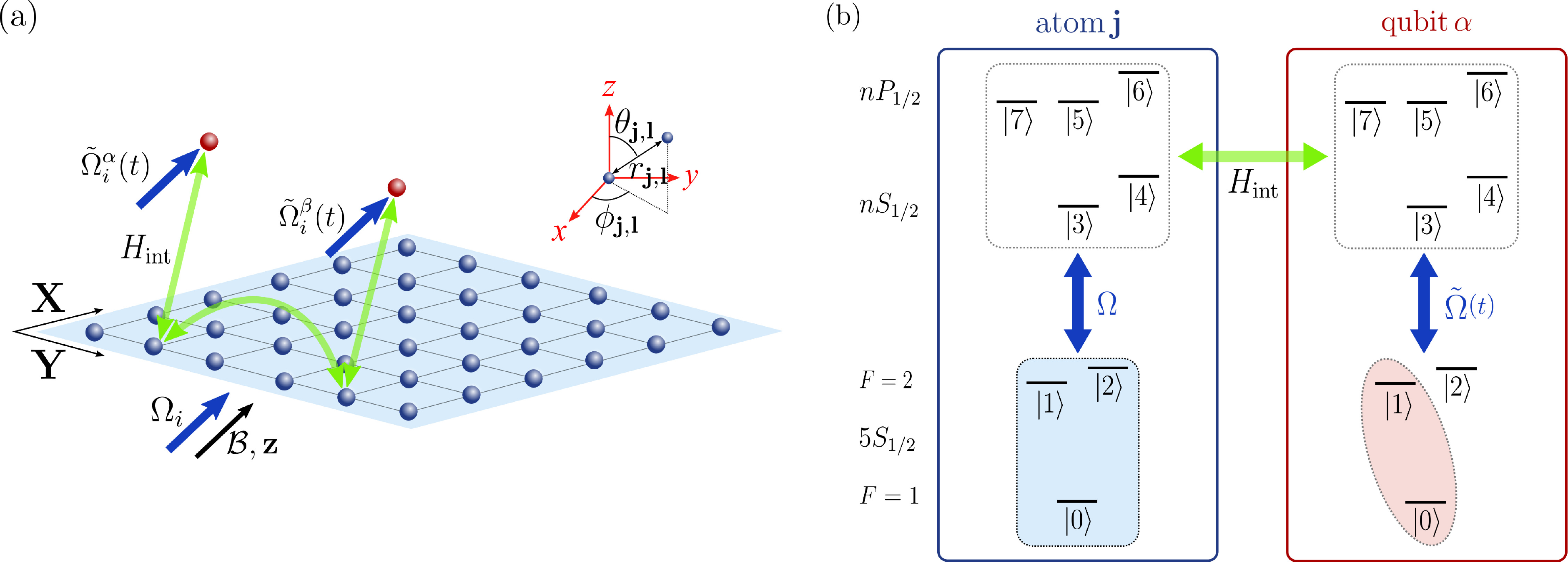}
\caption{Implementation of our model via Rydberg dressing.
(a) The qubits are placed in the vicinity of the TSS atoms, which are placed in a square lattice.
(b) The three-level structure of the topological spin system atoms is encoded in Zeeman states $\ket{0}$, $\ket{1}$, $\ket{2}$  of the hyperfine-structure ($F=1,2$ in the case of Rubidium atoms), so as the two levels $\ket{e}$, $\ket{g}$ of the qubits. We obtain an effective interaction between ground state levels by exciting them off-resonantly to Rydberg states where they interact via dipolar exchange interactions. \label{fig:setup}}
\end{figure}

Having introduced our model, we present an implementation using Rydberg-dressed ground-state atoms. 
Our proposal is based on the spin-orbit properties of the dipole-dipole interactions~\cite{Yao2012,Syzranov2014,Peter2015}.
We show schematically the different constituents of our implementation in Figure~\ref{fig:setup}(a) while the full level-structure and laser excitation scheme are detailed in \ref{app:atomic}.
The atoms which form our 2D topological spin system, denoted as TSS atoms, are trapped in a square lattice along the $X,Y$ plane as realized in optical lattices~\cite{Zeiher2016,Weber2015}, optical tweezers~\cite{Maller2015,Labuhn2016,Jau2015,Schlosser2012} or magnetic lattices~\cite{Leung2014,Herrera2015}.
We encode the states $\ket{0}$, $\ket{1}$, $\ket{2}$ in different hyperfine ground-states, for instance $F=1,2$ in the case of Rubidium atoms, where a magnetic field $\mathcal{B}\mathbf{z}$ with direction $\mathbf{z}=\sin\Theta \mathbf{X}+\cos\Theta\mathbf{Z}$ defines the quantization axis. The qubit atoms (red spheres) are placed in the vicinity of the topological spin system, using the same level structure as for the TSS atoms with $\ket{g}\equiv\ket{0}$ and $\ket{e}\equiv\ket{1}$. 
As depicted in Figure~\ref{fig:setup}(b) and written in detail in \ref{app:atomic} in the case of Rubidium atoms, TSS and qubit atoms are weakly excited by far-detuned laser fields to Rydberg states  $\lbrace\ket{3},\ket{4}\rbrace$, $\lbrace\ket{5},\ket{6},\ket{7}\rbrace$, which are magnetic Zeeman states of the $nS_{1/2}$ and $nP_{1/2}$ fine-structure manifolds, respectively.

We first present the effective Hamiltonian governing the dynamics of TSS ground-state atoms obtained by eliminating the Rydberg states in perturbation theory (Section~\ref{sec:perturbation}), and then apply the same approach for the qubits  (Section~\ref{sec:qubits}). Finally, we check the validity of our implementation by assessing the relevant time scales and the sources of imperfections (Section~\ref{sec:numbers}).

\subsection{Topological spin system}\label{sec:perturbation}

The TSS Hamiltonian $H_{\mathrm{T},\mathrm{imp}}=H_\mathrm{at}+ H_\mathrm{int}$ has two contributions. The first term $H_\mathrm{at}$ representing the energies of the levels and the laser excitation can be written in a rotating frame determined by the laser frequencies:
\begin{eqnarray}
H_\mathrm{at}&=&-\sum_{3\le i \le7} \Delta_i|i\rangle_\mathbf{j}\langle i| \nonumber
\\ &&+\sum_\mathbf{j} \left\lbrace   \Omega_0  \opdu{3}{\mathbf{j}}{0}+\Omega_1   \opdu{6}{\mathbf{j}}{1}+\Omega_2\left(\opdu{5}{\mathbf{j}}{2}+\frac{\sqrt{3}}{2}  \opdu{7}{\mathbf{j}}{1}\right) \nonumber\right.\\&&\quad\quad+\left.\mathrm{h.c.}\right\rbrace,\label{eq:Hat}
\end{eqnarray}
with the Rabi frequencies $\lbrace\Omega_0,\Omega_1\,\Omega_2\rbrace$, the laser detunings $\Delta_i$ and $\ket{i}_\mathbf{j}$  the state $i$ of atom $\mathbf{j}$. For simplicity, we consider in the following $\Delta_{i}=\Delta$.

As shown in \ref{app:atomic}, the second part of the Hamiltonian $H_\mathrm{int}$ representing the dipole-dipole interactions between Rydberg-excited states can be written as
\begin{eqnarray}
H_\mathrm{int}&=\sum_{\mathbf{j}\neq \mathbf{l}}  \frac{C_3}{r_\mathbf{j,l}^3} c^\dagger_\mathbf{j} h_\mathrm{dd}(\theta_\mathbf{j,l},\phi_\mathbf{j,l}) c_\mathbf{l}
\label{eq:Hint}
\end{eqnarray}
with the shorthand notation $c_\mathbf{j}=[|5\rangle_\mathbf{j}\langle 3|,|6\rangle_\mathbf{j}\langle 3|,|5\rangle_\mathbf{j}\langle 4|,|6\rangle_\mathbf{j}\langle 4|]$ and $h_\mathrm{dd}(\theta_\mathbf{j,l},\phi_\mathbf{j,l})$ a dimensionless $4\times4$ matrix, which depends on the spherical angles $\theta_\mathbf{j,l}$, and $\phi_\mathbf{j,l}$ of the relative vector $\mathbf{r}_\mathbf{j,l}$ with respect to $\mathbf{z}$. The radial coefficient  $C_3=\left(\int dr R_S(r) r^3 R_P(r)\right)^2$ is a function of $R_S$ ($R_P$), the radial wave function associated with the $S_{1/2}$ ($P_{1/2}$) Rydberg states~\cite{Saffman2010}.
We emphasize that we consider the large distance limit  $C_3/r_\mathbf{j,l}^3\ll\Delta$ where only the resonant flip-flop processes of the type $P_{1/2}S_{1/2}\to S_{1/2}P_{1/2}$ contribute.

In the perturbative (or dressing) regime, $\Omega_{0,1,2} \ll\Delta$, the Rydberg states can be eliminated in perturbation theory, leading to an effective Hamiltonian governing the dynamics of the ground state levels~\cite{Santos2000,Pupillo2010,Henkel2010}. To do so, we apply the Van Vleck formalism~\cite{Shavitt1980}  reducing the TSS Hamiltonian $H_{\mathrm{T},\mathrm{imp}}$ to the form of \eref{eq:HBreal}
with
\begin{eqnarray}
t_{\nu=1,2}(\mathbf{r_{j,l}})&=&(-1)^{\nu}\frac{C_3}{r_\mathbf{j,l}^3}\frac{ \Omega_{0}^{2} \Omega_{\nu}^{2}}{9 \Delta^4} \left(1- 3\cos^{2}\theta_\mathbf{j,l}\right)\label{eq:coupling1} \\
w(\mathbf{r_{j,l}})&=&\frac{C_{3}}{r_\mathbf{{j,l}}^3}\frac{ \Omega_{0}^{2} \Omega_{1} \Omega_{2}}{6 \Delta^{4}}\sin{\left (2 \theta_\mathbf{j,l} \right )},
\label{eq:coupling2}
\end{eqnarray}
and the second and fourth-order AC-Stark shifts
\begin{eqnarray}
\delta_1 &=&\frac{3\Omega_{2}^{2}}{4\Delta} - \frac{9\Omega_{2}^{4}}{16\Delta^3} + \frac{\Omega_{1}^{2}}{\Delta}  - \frac{\Omega_{1}^{4}}{\Delta^{3}} -\frac{3\Omega_{1}^{2} \Omega_{2}^{2}}{2\Delta^3}-\frac{ \Omega_{0}^{2}}{\Delta}+\frac{ \Omega_{0}^{4}}{\Delta^3}\nonumber \\
\delta_2 &=& \frac{\Omega_{2}^{2}}{\Delta} - \frac{\Omega_{2}^{4}}{\Delta^{3}} -\frac{ \Omega_{0}^{2}}{\Delta}+\frac{ \Omega_{0}^{4}}{\Delta^3} \label{eq:acstark},
\end{eqnarray}
where we have set $\delta_0=0$.  In order to obtain a topological phase~\cite{Peter2015}, it is important that the energy splitting $\delta_1-\delta_2$ does not dominate over the flip-flop term \eref{eq:coupling2}. This condition can be achieved for example by choosing the ratio $\Omega_2/\Omega_0$, $\Omega_1/\Omega_0$ according to \eref{eq:acstark} in order to obtain $\delta_1=\delta_2=0$. 

Finally, we emphasize, that in our dressing implementation, the local and time-dependent control of the laser intensities $\Omega_{0,1,2}$ allows to effectively disconnect atoms from the rest of the topological spin system and therefore to dynamically reshape the edges [c.f.~Figure~\ref{fig:concept}(d)].

\subsection{Qubits and coupling to the topological spin system}\label{sec:qubits}
The implementation of the qubits is similar to the one of the TSS atoms, with the difference that the level $\ket{2}$ is energetically excluded from the dynamics (for instance via the second-order AC-Stark shifts). Following the same procedure as for the derivation of the TSS Hamiltonian, we obtain the coupling Hamiltonian in the form of \eref{eq:HqB_real} with
\begin{eqnarray}
g_1(\mathbf{r}_{\mathbf{j},\alpha},t)&=&\frac{C_3}{r_{\mathbf{j},\alpha}^3}\frac{ \tilde\Omega_{0,\alpha}(t)\tilde\Omega_{1,\alpha}(t)\Omega_{0} \Omega_{1}}{9 \Delta^4}\left(3\cos^{2}{\theta_{\mathbf{j},\alpha}}-1\right)\\
g_2(\mathbf{r}_{\mathbf{j},\alpha},t)&=&\frac{C_{3}}{r_{\mathbf{j},\alpha}^3}\frac{\tilde \Omega_{0,\alpha}(t) \tilde\Omega_{1,\alpha}(t)\Omega_{0}\Omega_{2} }{6 \Delta^{4}}\sin{\left (2 \theta_{\mathbf{j},\alpha} \right )}e^{i\phi_{\mathbf{j},\alpha}},
\end{eqnarray}
where $\tilde\Omega_{0,\alpha}(t),\tilde\Omega_{1,\alpha}(t)$ denote the local Rabi frequencies addressing the qubits.
Finally, considering as in the case of the TSS atoms, that the AC-Stark shifts can be compensated, we set for simplicity the qubit Hamiltonian \eref{eq:Hqubit} to zero.

\subsection{Time scales}\label{sec:numbers}
We conclude this section by assessing the regime of validity of our implementation calculating the relevant time scales.
For a lattice spacing $a=15\,\mu$m and TSS Rabi frequencies $\Omega_0=\Omega_2=2\pi\times8\,\mathrm{MHz}$, $\Omega_1=2\pi\times 4\,\mathrm{MHz}$, detuning $\Delta=2\pi\times 25\,\mathrm{MHz}$ exciting the TSS atoms to Rydberg states with principal quantum number $n=65$, the value of the dipole-dipole coefficient $C_3=2\pi\times 19\ \hbar$GHz$\mu\mathrm{m}^3$ leads to flip-flop interactions of the order of $1$ kHz, which is larger than the effective decoherence rate $\Gamma\sim (\Omega/\Delta)^2\Gamma_\mathrm{r}\sim 2\pi\times 0.3$ kHz, induced by the Rydberg state admixture~\cite{Santos2000,Pupillo2010,Henkel2010} of the ground state atoms ($\Gamma_\mathrm{r}$ is the typical decay rate of the $S$ and $P$ Rydberg states~\cite{Beterov2009}).
Note that the strength of the interactions can be increased up to several kHz simply by reducing the lattice spacing~\footnote{In this situation, we would need however to calculate the dressing interactions numerically as the condition $C_3/a^3\ll \Delta$ used to derive analytical expression is not satisfied.}.
Finally we emphasize that we consider ground state atoms, which in contrast to Rydberg atoms, remain trapped while experiencing the dipole-dipole interactions and therefore decoherence effects arising from the motion of the atoms are negligible~\cite{Macri2014}.

\section{Application to quantum state transfer }\label{sec:statetransfer}
 In the following, we utilize our implementation to achieve quantum state transfer between two distant qubits $\alpha=1$, $\beta=2$.
 To do so, we calculate numerically the edge state properties of the topological spin system, solve the quantum state transfer protocol dynamics governed by the total Hamiltonian $H$ [c.f.~\eref{eq:H}] and compare our numerical results to the ideal predictions [\eref{eq:gamma1}-\eref{eq:gamma2}].
 
\subsection{Couplings to the edge state channel}\label{sec:kycode}
\begin{figure}[h!]
\centering
\includegraphics[width=\textwidth]{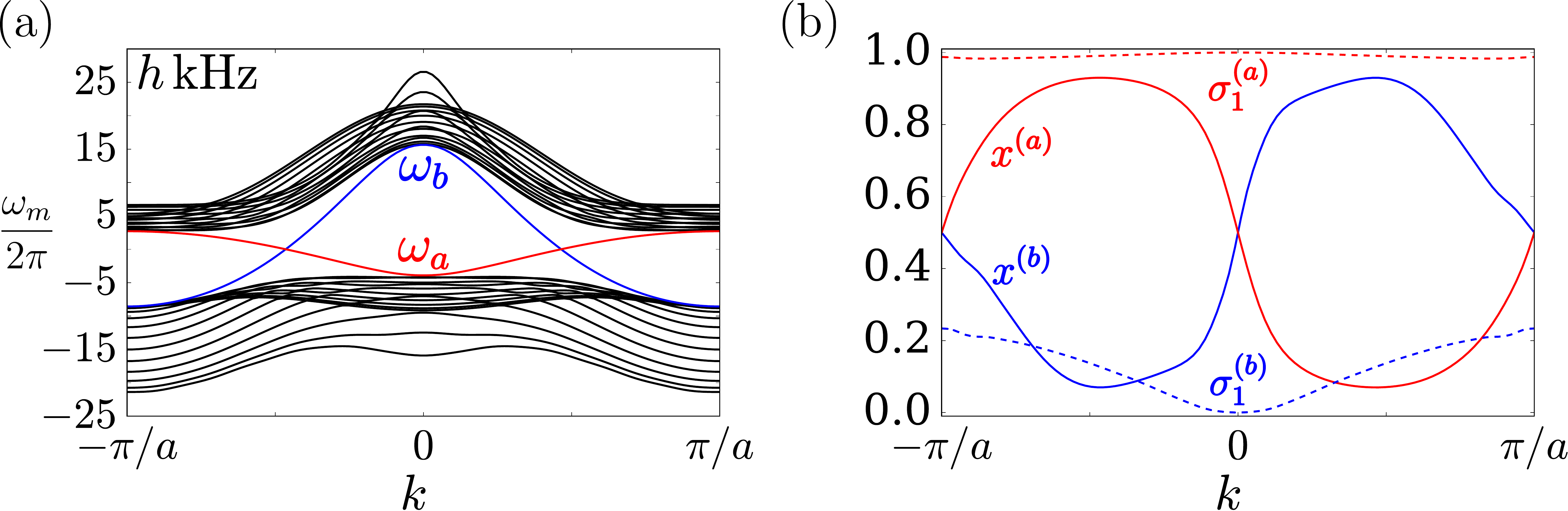}
\caption{Properties of the edge modes of the topological spin system. (a) Dispersion relation $\omega_m(k)$ obtained using the parameters given in Section~\ref{sec:numbers}, $\Theta=3\pi/4$, and $N_\mathrm{X} = 15$. The two edge states modes $m=\mathrm{a},\mathrm{b}$ are indicated as a blue and a red line.  (b)  Localization $x_\mathrm{a}(k)=(\sum_\nu(|c_{x=0,\nu}^{(k,\mathrm{a})}|^2-|c_{x=a(N_\mathrm{X}-1),\nu}^{(k,\mathrm{a})}|^2)+1)/2$ and spin $\sigma_{1,\mathrm{a}}(k)=\sum_{x}|c_{x,1}^{(k,\mathrm{a})}|^2$ of the two edge states which share the same chirality.  \label{fig:disp}}
\end{figure}
As shown in \emph{Peter et al.}~\cite{Peter2015} in the context of polar molecules, the TSS Hamiltonian exhibits a quantum Hall topological band structure phase associated with the existence of two edge modes $m=\mathrm{a},\mathrm{b}$.
We show in Figure~\ref{fig:disp} the dispersion relation $\omega_m(k)$ and the edge states localization and spin properties, obtained by numerically diagonalizing the TSS Hamiltonian using the dimension reduction analysis along the $Y$ axis (c.f.~\ref{app:kycode}) and for the numbers given in Section~\ref{sec:numbers}. The two edge state channels have the same chirality, i.e.~they propagate in the same direction along each edge. The existence of two edge modes offers in principle the advantage of performing multiplexing protocols but also gate operations based on spin-spin collisions (see~\cite{Ramos2016} in the context of spin chains). However, in the following we consider that one edge state mode $m=\mathrm{a}$ is predominantly excited, which can be achieved by exploiting the spin and localization properties of the edge states [c.f.~Figure~\ref{fig:disp}(b)].

\subsection{Quantum state transfer} \label{sec:numerics}
The quantum state transfer protocol relies on effective time-dependent couplings of the form of \eref{eq:pulse}, which in the  context of our implementation, can be achieved by dynamically varying the qubit Rabi frequencies $\tilde{\Omega}_{0,\alpha}(t),\tilde{\Omega}_{1,\alpha}(t)$ according to \eref{eq:gammaa}, \eref{eq:coupling1} and \eref{eq:coupling2}.
 We study the performance of the protocol by numerically calculating the corresponding dynamics governed by \eref{eq:H} with the initial condition $|\Psi(t=0)\rangle =  \sigma_1^+|V\rangle$.
As an example, we choose the two qubits to be separated by a distance $d=y_2-y_1$ and position them with respect to the closest TSS atom $\mathbf{j}[\alpha]$ (with $x_{\mathbf{j}[\alpha]}=0$) via $\mathbf{r}_{\mathbf{j}[\alpha],\alpha}=b(\mathbf{Z} -  \mathbf{X})$, with $b=8.1\,\mathrm{\mu m}$.
For this geometric configuration, we obtain a dominant coupling $\gamma_a$ to the edge state indicated in red in Figure~\ref{fig:disp}.
\begin{figure}[h]
\includegraphics[width=\columnwidth]{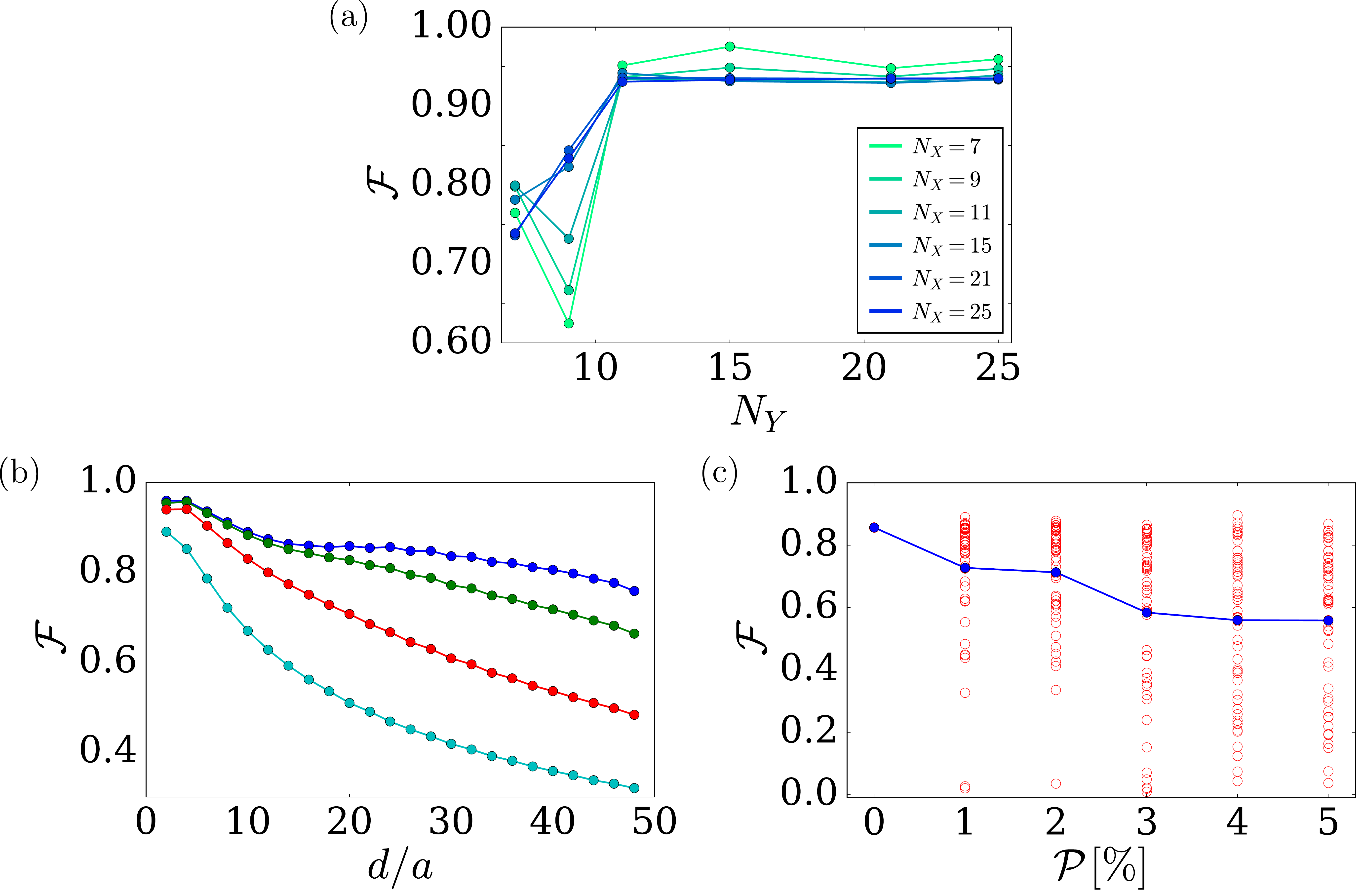}
\caption{Quantum state transfer fidelities $\mathcal{F}$ with $\Theta=3\pi/4$. The pulse parameters are $\gamma_0 t_0=6$ and $c =1.01\pi/4\gamma_{0}^{2}$. 
(a) Influence of the size of the topological spin system with $d=6a$ and $\gamma_0=2\pi\times0.64\,\mathrm{kHz}$.
(b)  Fidelity as a function of $d$ for coupling strengths  $\gamma_0=2\pi\times0.64\,\mathrm{kHz}$ (blue), $\gamma_0=2\pi\times0.8\,\mathrm{kHz}$ (green), $\gamma_0=2\pi\times1.11\,\mathrm{kHz}$ (red) and $\gamma_0=2\pi\times1.60\,\mathrm{kHz}$ (cyan), showing the importance of dispersive effects at large distances. The TSS extension is set by $N_\mathrm{X}=15$, $N_\mathrm{Y}=61$
(c) Fidelity as a function of the probability of site vacancies showing the destructive impact of disorder on the efficiency of the quantum state transfer, for $N_\mathrm{X}=15$, $N_\mathrm{Y}=41$.
 \label{fig:results}}
\end{figure}

The results of the quantum state transfer protocol are shown in Figure~\ref{fig:results} where we represent the fidelity of the quantum state transfer $\mathcal{F}=|c_{e,2}(t_f)|^2$ for a final time $t_f=8$ ms. In panel (a), we study the effect of the finite size of the topological spin system by representing the fidelity $\mathcal{F}$ for different TSS sizes $(N_\mathrm{X},N_\mathrm{Y})$, a distance $d=6a$ and $\gamma_0=2\pi\times0.64\,\mathrm{kHz}$.
For large TSS sizes $N_\mathrm{X},N_\mathrm{Y}>10$ the fidelity converges towards a constant value $\mathcal{F}\approx0.95$ showing that the dispersion relation, which we derived under the assumption of a semi-infinite topological spin system, is relevant to describe the quantum state transfer in large but finite systems.
In order to explain the cause of the deviation from the ideal fidelity $\mathcal{F}=1$, we represent $\mathcal{F}$ as a function of $d$ and $\gamma_0$  in panel (b).
At short distances and small coupling strengths, we observe state transfer with a small error which we attribute to the influence of the second edge state channel $m=\mathrm{b}$ and to the off-resonant contribution of the bulk modes $m\neq\mathrm{a},\mathrm{b}$.  At large distances, the fidelity is a decreasing function of $d$ and of the coupling strength $\gamma_0$ indicating the onset of dispersion effects~\cite{Ramos2016} which distort the edge state wave-packet while propagating.

Furthermore, we study the influence of disorder, which in cold atom experiments typically manifests by an on-site probability $\mathcal{P}$ for TSS atom vacancies.
The fidelity $\mathcal{F}$ as a function of $\mathcal{P}$ is shown in Figure~\ref{fig:results}(b) for a fixed distance $d=26a$ and coupling $\gamma_0=2\pi\times0.64\,\mathrm{kHz}$.
Despite the absence of back-scattering in our topological quantum channel, the presence of disorder distorts wave-packets of edge states, affecting the fidelity of the quantum state transfer.
Moreover, considering the individual disorder realizations shown as red circles, we notice that the robustness of the protocol crucially depends on the position of the vacancies.

To summarize this section, we have demonstrated the potential of using a Rydberg dressing implementation of chiral channels for quantum state transfer. We studied the role of dispersion and disorder as sources of imperfection, the latter having a crucial influence on the fidelity despite the topological character of the edge modes. In the next section, we show that these limitations can be overcome by performing a protocol based on the spectroscopy of the transmission channel.

\section{A robust state transfer protocol for the effects of dispersion and disorder}\label{sec:chirp}

The goal of this section is to present a state transfer protocol which is robust against dispersive and disorder effects.
In contrast to the standard protocol which assumes that the wave-packet propagates without deformation according to a linear dispersion relation, here we simply treat the quantum channel as a ``black box'' which conveys the information from the first to the second qubit.
This section is organized as follows. In Section~\ref{sec:spectro} we show how to realize spectroscopy of the quantum channel via the qubits, i.e.~to measure the phase and amplitude of the distorted wave-packet. We then use this information to derive in Section~\ref{sec:pulses} the protocol which achieves the quantum state transfer. Finally, in Section~\ref{sec:toy}, we illustrate  the efficiency of our method using two toy models.
\begin{figure}[h!]
\includegraphics[width=0.85\columnwidth]{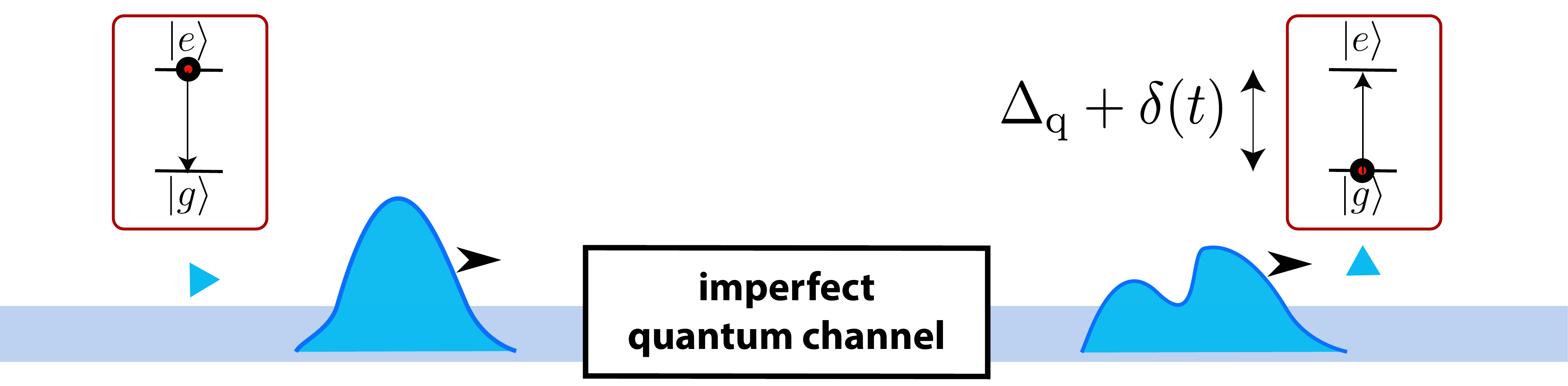}
\caption{Robust state transfer with static imperfections: the first qubit emits a symmetric wave-packet towards the second qubit. Its unknown final shape can be measured by the second qubit using a chirped transition frequency $\Delta_\mathrm{q}+\delta(t)$. We use this information to derive the pulse shape $\gamma_{\mathrm{a},2}(t)$ and chirp $\delta$ realizing the quantum state transfer. \label{fig:spectro}}
\end{figure}

We emphasize that our protocol is based on the assumption of perfect chirality of the quantum channel, i.e.~relies on the fact that all the emission of the first qubit is transmitted towards the second qubit.
It thus applies in the context of the topological spin system presented in this previous sections but also in the context of chiral 1D waveguides~\cite{Ramos2016} subject to dispersive effects~\footnote{In this case however, the channel is not protected against back-scattering.}.

\subsection{Quantum channel spectroscopy using the qubits}\label{sec:spectro}

We now present the different steps of the spectroscopy of the quantum channel. As depicted in Figure~\ref{fig:spectro}, we consider the first qubit $\alpha=1$ to be initially excited and to emit for times $t<0$ a wave-packet with a well-defined shape which is then distorted while propagating. At time $t=0$, the wave-packet reaches the second qubit, whose dynamics, following \ref{app:WW},  is given by 
 \begin{eqnarray}
\dot c_{e,2}(t) &=& -\left[\frac{\gamma_{\mathrm{a},2}(t)}{2}+i\delta(t)\right] c_{e,2}(t)+\sqrt{\gamma_{\mathrm{a},2}} f(t) \label{eq:c},
 \end{eqnarray}
with $f(t)=-i v_\mathrm{a}(\bar k_\mathrm{a})  e^{i \bar k_\mathrm{a} v_\mathrm{a}(\bar k_\mathrm{a}) t -i \eta_2} c_{y_2-v_\mathrm{a}(\bar k_\mathrm{a})t}(0)$ where $c_{y}(t)=\int dk e^{iky} c_{k,\mathrm{a}}(t)$ represents the wave-packet amplitude in real space and we assume $c_{e,1}(0)\approx 0$.
In contrast to the standard state transfer protocol which supposes that a wave-packet propagates without deformation~\cite{Cirac1997}, we assume here that the second qubit receives an unknown wave-packet $f(t)$ whose amplitude $|f(t)|$ and phase $\Phi(t)\equiv \mathrm{arg}(f(t))$~\footnote{originated for instance from dispersion effects.} are both unknown~\footnote{Note that the assumption of perfect chirality of the quantum channel implies $\int dt |f(t)|^2 dt = 1$.}. 
The time-dependent quantity $\delta(t)$ (c.f.~Figure~\ref{fig:spectro}) allows to change the transition frequency of the second qubit dynamically, which is assumed not to modify the coupling $\gamma_{\mathrm{a},2}(t)$. As  explained in Section~\ref{sec:pulses} the chirp $\delta(t)$ is a crucial ingredient to realize the quantum state transfer.

We now show how to measure the wave-packet $f(t)$. Using the ansatz $d(t)=e^{\gamma_{\mathrm{a},2}t/2}c_{e,2}(t)$, \eref{eq:c} becomes
\begin{equation}
 \dot d = -i\delta d +\sqrt{\gamma_{\mathrm{a},2}} e^{\gamma_{\mathrm{a},2}t/2} f(t) \label{eq:d}.
 \end{equation}
In a typical experimental setup, we only have access to the population of the qubit, and thus to the modulus $r\equiv|d|$, while the phase $\theta\equiv\mathrm{arg}(d)$ cannot be directly measured. Therefore, it is not possible to extract the function $f(t)$ via \eref{eq:d} with a single measurement of the qubit population.
However, the measurement can be repeated for different (time-independent) $\delta$ leading to the response $d(t)$ and consequently $f(t)$.
A simple option is to measure $r_0$ for $\delta=0$ and $r'_0\equiv (\partial r/\partial \delta)_{\delta=0} $, i.e.~its derivative with respect to $\delta$, at $\delta=0$. We can then  eliminate $f(t)$ using \eref{eq:d} and obtain
 \begin{eqnarray}
 \dot r'_0-\dot \theta_0 u&=&0 \label{eq:sys1}\\
 \dot \theta_0 r'_0+\dot u +r_0 &=&0 \label{eq:sys2},
 \end{eqnarray}
 with $\theta'_0\equiv (\partial \theta/\partial \delta)_{\delta=0}$, $u=r_0 \theta'_0$.
 Knowing $r_0$ and $r_0'$, \eref{eq:sys1},\eref{eq:sys2} can be integrated to find $\theta_0$ and finally $f(t)$ using \eref{eq:d}. 
 As we are interested here in studying deviations from the ideal case where the function $f(t)$ is real, we solve the differential equations \eref{eq:sys1},\eref{eq:sys2} iteratively in $\dot\theta_0$, the zeroth order $\dot \theta_0(t) = -\dot r'_0(t)/\int r_0(t') dt'$ will be in the examples below a good approximation. This concludes the spectroscopy of the quantum channel: comparing $f(t)$ with the initial pulse sent by the first qubit indicates how a wave-packet is transmitted towards the second qubit including the effects of dispersion and disorder. We now use this information to realize a robust state transfer protocol.
 
 \subsection{Pulse shapes}\label{sec:pulses}
 
Under the assumption of static quantum channel imperfections, the knowledge of the wave-packet $f(t)$, which reaches the second qubit, can be used to realize a robust state transfer protocol. For a perfect absorption of the entire wave-packet $f(t)$ with the evolution $c_{e,2}(t)=e^{i\Phi(t)}\sqrt{\int_0^t |f(t')|^2 dt'}$, we obtain the conditions
\begin{eqnarray}
 \gamma_{\mathrm{a},2}(t)&=&|f(t)|^2/\int_0^t |f(t')|^2 dt' \label{eq:cond1}\\
 \delta(t)&=&-\dot \Phi(t) \label{eq:cond2}, 
 \end{eqnarray} 
where we used \eref{eq:c}. The first condition \eref{eq:cond1} resembles the typical pulse shape obtained in the standard state transfer protocol~\cite{Cirac1997,Gorshkov2007b,Stannigel2011} with a real envelop $f(t)$ [c.f.~\eref{eq:gamma2} in the Gaussian case]. The second ``phase-matching ''condition allows to compensate for the existence of the phase $\Phi$ by a ``chirped'' frequency $\delta(t)$. In this way, the frequency of the second qubit is dynamically synchronized with the evolution of the phase $\Phi(t)$.

\subsection{Results}\label{sec:toy}
We now apply our protocol based on the spectroscopy of the channel, using two toy models. First, in Section~\ref{sec:1d}, we consider a 1D quantum channel subject to dispersive effects while we present in Section~\ref{sec:2d} the results obtained in the case of a disordered topological spin system. In both models, we assess the efficiency of the protocol by numerically simulating the dynamics of the combined system formed by the qubits and the quantum channel, similar to the study presented in Section~\ref{sec:statetransfer}.

\subsubsection{Compensation of dispersion effects in structured 1D waveguides}\label{sec:1d}
\ \\
We consider a unidimensional waveguide ($N_\mathrm{X}=1$), where the excitations are encoded in a single excited state $|1\rangle$. The matrix $h(\mathbf{r_{j,l}})$ [c.f.~\eref{eq:HBreal}] is a simple scalar with nearest neighbor interactions $h(\mathbf{r_{j,l}})=-J\delta_{j_y,l_y\pm 1}$. This model has been studied in detail in~\cite{Ramos2016} showing dispersive effects, similar to the ones presented in Section~\ref{sec:statetransfer}. In this case, the dispersion relation has an analytical expression $\omega_\mathrm{a}(k)=-2J\cos (ka)+\delta_\nu$ where we choose $\delta_\nu=0.5 J$, $\Delta_\mathrm{q}=0$.
\begin{figure}[h]
\includegraphics[width=0.978\columnwidth]{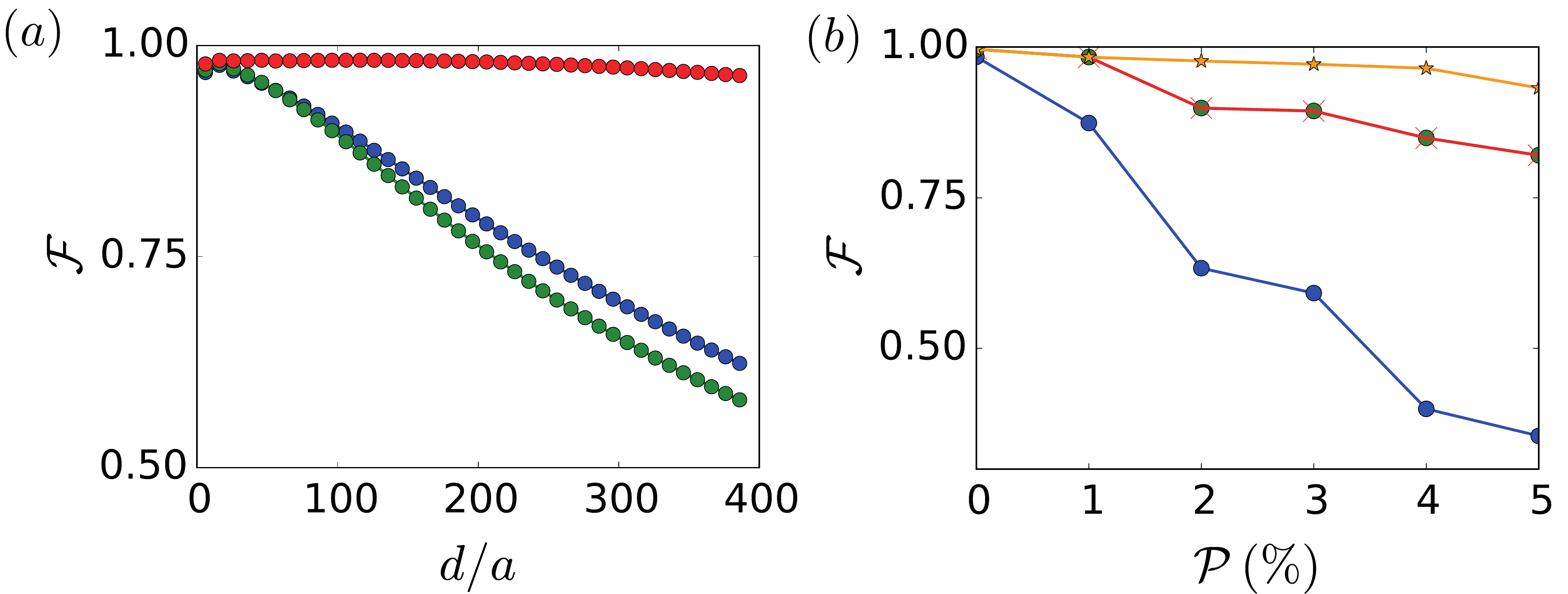}
\caption{Correction of (a) dispersion effects in the case of a spin chain and of (b) disorder with a topological spin system. In each case the first qubit emits a Gaussian wave-packet according to \eref{eq:pulse} with $\gamma_0=0.3J$. In panel (b), the fidelity has been averaged over 50 realizations of the disorder. We represent in blue the standard state transfer protocol [\eref{eq:gamma1},\eref{eq:gamma2}], in green the correction of the pulse shape and in red the case where the chirp is also applied. In the case of the spin chain (a), the chirp is an essential ingredient to compensate for dispersion effects. With disorder and a topological spin system, green and red curved overlap showing that the chirp is not needed. Finally, the orange curve representing the averaged fidelity where the coupling sites $\mathbf{j}[\alpha]$ are not affected by the disorder shows that the protocol only applies when the disorder does not affect the couplings $\gamma_{\mathrm{a},\alpha}$ of the qubits to the waveguide. \label{fig:results_chirp}}
\end{figure}
The fidelity of the quantum state transfer protocol is shown in Figure~\ref{fig:results_chirp}(a). The blue curve corresponds to the standard state transfer protocol, showing how dispersion affects the quantum state transfer at large distances $d/a$. The green curve represents the case where the coupling of the second qubit has been adapted  according to \eref{eq:cond1} whereas the red curve also includes the ``chirp'' condition \eref{eq:cond2}. The second condition is the crucial requirement to obtain a robust state transfer protocol. For a distance $d=400a$, we obtain a fidelity of 96\%, corresponding to a reduction of the error of $90\%$  compared to the standard state transfer protocol. 
The robustness of the protocol simply comes from the fact that the chirp compensates for the phases accumulated by the Fourier components of the propagating wave-packet. 

\subsubsection{Robustness against disorder in a topological spin system}\label{sec:2d}
\ \\ 
Finally, we study the robustness of the protocol in the case of a disordered topological spin system. For simplicity, we consider a model which in contrast to the Rydberg implementation Section~\ref{sec:implementation} includes a single edge state channel~\cite{Qi2008}. The Hamiltonian is commonly written in terms of Pauli matrices
\begin{eqnarray}
H_\mathrm{T}&=&\sum_\mathbf{j} J\left(b^\dagger_{j_x,j_y}\frac{\sigma_z-i\sigma_x}{2} b_{j_x+1,j_y}  +b^\dagger_{j_x,j_y} \frac{\sigma_z-i\sigma_y}{2} b_{j_x,j_y+1}+\mathrm{h.c.}\right) \nonumber\\
			&+&m\sum_\mathbf{j} b^\dagger_\mathbf{j} \sigma_z b_\mathrm{\mathbf{j}}, 
\end{eqnarray}
where we fix in the following $m=-1$. Moreover, the two qubits are coupled to a single TSS atom $\mathbf{j}[\alpha]$ (with $y_{\mathbf{j}[\alpha]}=0$) according to \eref{eq:HqB_real} with $g_\nu(\mathbf{r}_{\mathbf{j},\alpha},t)=-\tilde J(t) (-1)^{\nu}/\sqrt{2}\delta_{\mathbf{j},\mathbf{j}[\alpha]}$.

The results of the protocol are shown in Figure~\ref{fig:results_chirp}(b) with the same graphical conventions as in Figure~\ref{fig:results_chirp}(a).
The comparison between the different curves shows that adapting the coupling pulse according to \eref{eq:cond1} increases substantially the fidelity $\mathcal{F}$ of the quantum state transfer, while adding a chirp term [c.f.~\eref{eq:cond2}] is in this case not required.
Our interpretation is that in contrast to dispersion effects, the effect of missing atoms mainly leads to a time-delayed absorption of the wave-packet, corresponding to the time needed by the wave-packet to ``avoid'' the defects.
Finally, we note that our protocol is not robust when the coupling between the qubits and the edge state channel is affected by the disorder, which in this example occurs when one site $\mathbf{j}[\alpha]$ is vacant. This is illustrated in Figure~\ref{fig:results_chirp}(b) by the orange line, which represents the averaged fidelity for disorder realizations where the two sites $\mathbf{j}[1],\mathbf{j}[2]$ are both occupied and corresponds to a much higher fidelity compared to the red curve.

\section{Conclusion}\label{sec:conclusion}
In summary, we have studied a model of a quantum network where qubits can interact via chiral edge states. Our implementation based on Rydberg-dressed ground state atoms allows to demonstrate the different ingredients of a quantum state transfer using a topologically protected spin system in state of the art experimental setups, and can be easily adapted to other dipolar systems such as polar molecules.
Furthermore, after having numerically studied the role of static imperfections in the standard protocol [\eref{eq:gamma1},\eref{eq:gamma2}], we have presented an original approach, based on the spectroscopy of the quantum channel, achieving high-fidelity quantum state transfers even in the presence of dispersive and disorder effects.

In a broader context, our model of chiral quantum network in dipolar arrays can be applied to realize various robust quantum operations using the chiral edge states as topologically protected quantum channels.
With directional spin chains, the hard-core nature of spin excitations makes it possible to implement entangling  gates between distant qubits~\cite{Ramos2016,Gorshkov2010}.
In the case of dipolar topological spin systems, where the equilibrium phase diagram includes a Fractional Chern insulator~\cite{Yao2013}, the role of topology in the collision dynamics of multiple edge state excitations and the opportunities for realizing entangling gates represent fundamental questions for quantum information processing in quantum networks, which we plan to address in a future work.
\ack
We thank H. Ter\c{c}as, M. Dalmonte, Y.Hu, J.Budich and T.Ramos for useful discussions. The numerical solutions of the Schr\"odinger equation were obtained using the QuTiP toolbox~\cite{Johansson20131234}. Work at Innsbruck is supported by the EU (UQUAM, SIQS, RYSQ), the SFB FOQUS of the Austrian Science Fund. and the Army Research Laboratory Center for Distributed Quantum Information via the project SciNet

\section*{References}
\bibliography{paper}
\appendix
\section{Diagonalization of the TSS Hamiltonian}\label{app:kycode}
In this section we show how to obtain the dispersion relation describing the topological spin system. To do so, we consider the system to be infinite in the $Y$ direction, while remaining finite in the $X$ direction and diagonalize the TSS Hamiltonian ~\eref{eq:HBreal}. 
The presence of at most one excitation in the TSS, $\sum_{j,\nu} \langle b_{\mathbf{j},\nu}^\dagger b_{\mathbf{j},\nu} \rangle \le 1$, allows us to treat the hard-core boson operators  $b_{\mathbf{j},\nu}$ as genuine bosonic operators. Using the transformation $b_{\mathbf{j},\nu}=(1/\sqrt{2\pi})\int_{-\pi/a}^{\pi/a}\mathrm{d}k\,e^{iky_\mathbf{j}} b_{x_\mathbf{j},k,\nu}$, we obtain
\begin{equation}
H_\mathrm{T} =\sum_{x_\mathbf{j},x_\mathbf{l}}\int\mathrm{d}k\mathbf{b}^\dagger_{x_\mathbf{j},k}  \tilde h(x_\mathbf{j}-x_\mathbf{l},k) \mathbf{b}_{x_\mathbf{l},k}
\label{eq:HBk}
\end{equation}
where $\mathbf{b}_{x_\mathbf{l},k}=[b_{x_\mathbf{l},k,0},b_{x_\mathbf{l},k,1}]$, $k$ is the wave-vector associated with a plane-wave moving along the $Y$ direction and 
\begin{equation}
 \tilde h(x_\mathbf{j}-x_\mathbf{l},k)= \sum_{y} e^{-ik y}h((x_\mathbf{j}-x_\mathbf{l})\mathbf{X}+y\mathbf{Y}),
\end{equation}
with $h(\mathbf{0})=0$. Finally, the TSS Hamiltonian \eref{eq:HBk} can be written in the quadratic form of \eref{eq:HBk1} using the operators $b_{k,m}=\sum_{x,\nu}c_{x,\nu}^{(k,m)}b_{x,k,\nu}$ representing the eigenmodes of the Hamiltonian and $\omega_m(k)$ the corresponding dispersion relation. 

\section{Wigner-Weisskopf treatment of qubits coupled to edge-modes\label{app:WW}}
In this section we consider the model introduced in Section~\ref{sec:model} to derive general expressions for the dynamics of the qubits (c.f. \eref{eq:gamma1},\eref{eq:gamma2}). Starting from the Wigner Weisskopf ansatz~\eqref{eq:WW}
and plugging it into the Schr\"odinger Equation $d\ket{\psi(t)}/dt=-i(H_\mathrm{q}+H_\mathrm{T}+H_\mathrm{qT})\ket{\psi(t)}$ leads to a set of coupled differential equations for the amplitudes
\begin{eqnarray}
\dot{c}_{e,\alpha}(t)&=& -i\sum_m \int\mathrm{d}k\,(g_{k,m}^{(\alpha)}(t))^* c_{k,m}(t) e^{i k y_\alpha}e^{-i(\omega_m(k)-\Delta_\mathrm{q}) t}\label{eq:ampl1}\\
\dot{c}_{k,m}(t)&=&-i\sum_\alpha g_{k,m}^{(\alpha)}(t) c_{e,\alpha}(t) e^{-i ky_\alpha}e^{i(\omega_m(k)-\Delta_\mathrm{q}) t}\label{eq:ampl2}.
\end{eqnarray}
Formal integration of \eref{eq:ampl2} 
\begin{equation}
c_{k,m}(t)=-i\sum_\alpha\int_0^t \mathrm{d}t'\,g_{k,m}^{(\alpha)}(t') c_{e,\alpha}(t') e^{-i k y_\alpha}e^{i(\omega_m(k)-\Delta_\mathrm{q}) t'}\label{eq:ampl3}
\end{equation}
and \eref{eq:ampl3} plugging into \eref{eq:ampl1} gives with $y_{\alpha,\beta}\equiv y_\alpha-y_\beta$
\begin{equation}
\dot{c}_{e,\alpha}(t)=-\sum_{m,\beta}\int_0^t\mathrm{d}t'\int\mathrm{d}k\,(g_{k,m}^{(\alpha)}(t))^* g_{k,m}^{(\beta)}(t') c_{e,\beta}(t') e^{iky_{\alpha,\beta}} e^{-i(\omega_m(k)-\Delta_q)(t- t')}.
\end{equation}
If we assume that the qubit timescales are slow compared to the bath timescales (weak-coupling regime $h(\mathbf{r_{j,l}})\gg g_\nu(\mathbf{r}_{j,\alpha})$) we can linearize the dispersion relation for a particular edge-mode $\omega_\mathrm{a}(k)$ around the qubit resonance $\omega_\mathrm{a}(\bar{k}_\mathrm{a})=\Delta_\mathrm{q}$
\begin{equation}
\omega_\mathrm{a}(k)\approx\Delta_\mathrm{q}+(k-\bar{k}_\mathrm{a})v_\mathrm{a}(\bar{k}_\mathrm{a}).
\end{equation}
Furthermore, the weak coupling approximation allows us to assume that $g_{k,\mathrm{a}}^{(\alpha)}(t)$ is independent of $k$ around the resonant wavevector $\bar{k}_\mathrm{a}$, such that $g_{k,\mathrm{a}}^{(\alpha)}(t)\approx g_{\bar{k}_\mathrm{a},\mathrm{a}}^{(\alpha)}(t)$. This leads to the following expressions for the qubit amplitudes
\begin{eqnarray}
\dot{c}_{e,\alpha}(t)&=&-\sum_{\beta}\int_0^t\mathrm{d}t' \sqrt{\gamma_{\mathrm{a},\alpha}(t)\gamma_{\mathrm{a},\beta}(t')}e^{i\lbrace\bar{k}_\mathrm{a}y_{\alpha,\beta}-\eta_{\alpha,\beta}\rbrace}c_{e,\beta}(t')\nonumber\\&&\quad \times\delta\left(t-y_{\alpha,\beta}/v_\mathrm{a}(\bar{k}_\mathrm{a})-t'\right)\\
&=&-\frac{1}{2}\gamma_{\mathrm{a},\alpha}(t) c_{e,\alpha}(t)\nonumber\\
&&-\sum_{\beta\neq\alpha}\Theta(y_{\alpha,\beta}/v_\mathrm{a}(\bar{k}_\mathrm{a}))
\Theta(t-y_{\alpha,\beta}/v_\mathrm{a}(\bar{k}_\mathrm{a}))
e^{i(\bar{k}_\mathrm{a}y_{\alpha,\beta}-\eta_{\alpha,\beta})}\nonumber\\&&\quad\times \sqrt{\gamma_{\mathrm{a},\alpha}(t)\gamma_{\mathrm{a},\beta}(t-y_{\alpha,\beta}/v_\mathrm{a}(\bar{k}_\mathrm{a}))}c_{e,\beta}(t-y_{\alpha,\beta}/v_\mathrm{a}(\bar{k}_\mathrm{a})) 
\label{cef}\end{eqnarray} 
with $\gamma_{\mathrm{a},\alpha}(t)=(2\pi|g_{\bar{k}_\mathrm{a},\mathrm{a}}^{(\alpha)}(t))|^2)/|v_\mathrm{a}(\bar{k}_\mathrm{a})|$, $g_{k,\mathrm{a}}^{(\alpha)}(t)=|g_{k,\mathrm{a}}^{(\alpha)}(t)|e^{i\eta_\alpha}$, $\eta_{\alpha,\beta}\equiv\eta_\alpha-\eta_\beta$ and the Heaviside function $\Theta(x)$ defined as $\Theta(x)=1$ for $x > 0$  and  $\Theta(x)=0$ for $x\leq 0$.

\section{Dressing scheme details\label{app:atomic}}
\begin{figure}[h!]
\includegraphics[width=\textwidth]{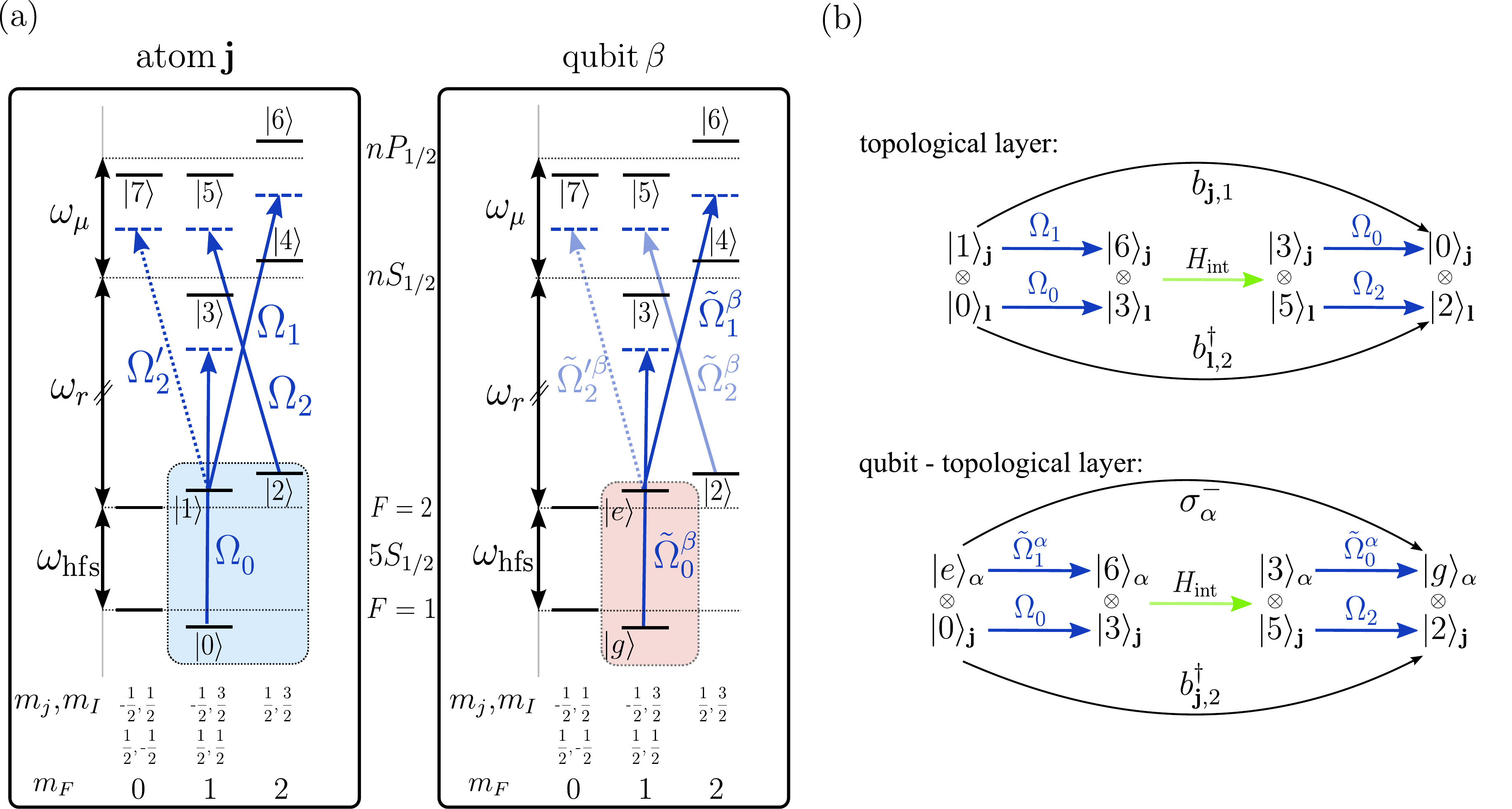}
\caption{$(\mathrm{a})$ Dressing scheme used to implement the interaction terms of our model [c.f.~Section~\ref{sec:model}] $(\mathrm{b})$ Example for excitation transfer process within the topological spin system and between qubit and TSS using the dressing scheme. \label{fig:app}}
\end{figure}

In this section we present the details of our dressing scheme. The atomic levels we are interested in are represented in Figure~\ref{fig:app}(a) in the context of Rubidium atoms. In particular the Rubidium hyperfine ground states $\ket{0}\equiv|5S_{1/2},F=1,m_F=1\rangle$ and $\ket{1}\equiv|5S_{1/2},F=2,m_F=1\rangle$, $\ket{2}\equiv|5S_{1/2},F=2,m_F=2\rangle$ represent the vacuum state and the two excited states of our model.

They are excited off-resonantly to the Rydberg states $\ket{3}\equiv|nS_{1/2},m_j=-1/2\rangle\otimes|m_I=3/2\rangle$, $\ket{4}\equiv|nS_{1/2},m_j=1/2\rangle\otimes|m_I=3/2\rangle$,  $\ket{5}\equiv|nP_{1/2},m_j=-1/2\rangle\otimes|m_I=3/2\rangle$, $\ket{6}\equiv|nP_{1/2},m_j=1/2\rangle\otimes|m_I=3/2\rangle$ and $\ket{7}\equiv|nP_{1/2},m_j=-1/2\rangle\otimes|m_I=1/2\rangle$~\footnote{It is important to note that, for the time scales of our implementation, the hyperfine interaction between Rydberg states is negligible so that the nuclear spin $m_I$ behaves as a spectator in the dynamics.}, which will be used to generate the hopping Hamiltonian \eref{eq:HBreal} using three laser beams propagating along the $z$ direction with optical frequencies $\omega_0,\omega_1,\omega_2$ (where the laser frequency $\omega_0$ is associated with a two-photon process), with respectively linear, $\sigma_+$ and $\sigma_-$ polarization. We emphasize that the state $\ket{7}$ has a different nuclear moment and will only lead to an AC-Stark shift contribution.
The frequencies $\omega_\mathrm{hfs}$, $\omega_{r}$ and $\omega_{\mu}$ [c.f.~Figure~\ref{fig:app}(a)] denote respectively the hyperfine splitting in the ground state, the energy separation between ground and Rydberg states and the energy difference between the two Rydberg manifolds $nS_{1/2}$ and $nP_{1/2}$. According to the Land\'e factor and the strength $\mathcal{B}$ of the magnetic field, we further obtain the Zeeman shifts $\epsilon_0=-1/2 \mu_B \mathcal{B}$, $\epsilon_1=1/2 \mu_B \mathcal{B}$, $\epsilon_2= \mu_B \mathcal{B}$, $\epsilon_3= -\mu_B \mathcal{B}$, $\epsilon_4= \mu_B \mathcal{B}$, $\epsilon_5=-1/3 \mu_B \mathcal{B}=\epsilon_7$, $\epsilon_6=1/3 \mu_B \mathcal{B}$ ($\mu_B$ is the bohr Magneton). At this point it is important to note, that one needs to fulfill the condition $\omega_1+\epsilon_1=\omega_2+\epsilon_2$ in order to keep the interaction Hamiltonian~\eqref{eq:Hint} time-independent. In the frame rotating with the laser frequencies the detunings appearing in \eref{eq:Hat} are given by $\Delta_i=\omega_0-\omega_\mathrm{hfs}-\omega_r-\epsilon_i+\epsilon_0$ ($i=3,4$) and $\Delta_5=\omega_2-\omega_r-\omega_\mu-\epsilon_5+\epsilon_2$,  $\Delta_6=\omega_1-\omega_r-\omega_\mu-\epsilon_6+\epsilon_1$, $\Delta_7=\omega_2-\omega_r-\omega_\mu-\epsilon_5+\epsilon_1$. For simplicity it is assumed that the Zeeman shifts are negligible compared to the laser detunings. Finally the quantity $\Omega_2'/\Omega_2=\sqrt{3}/2$ is given by the ratio of the dipole matrix elements between the involved hyperfine ground states and the Rydberg levels.

The second part of the Hamiltonian~\eref{eq:Hat} represents the dipole-dipole interaction between two atoms in Rydberg states. Two atoms $\mathbf{j}$, $\mathbf{l}$ interact at long distances via the dipole-dipole potential~\cite{Saffman2010}
\begin{equation}
V_{\mathrm{dd}}^{(\mathbf{j,l})}=\frac{\hat{\mathrm{d}}^{(\mathbf{j})}\hat{\mathrm{d}}^{(\mathbf{l})}-3(\hat{\mathrm{d}}^{(\mathbf{j})}\hat{\mathrm{n}})(\hat{\mathrm{d}}^{(\mathbf{l})}\hat{\mathrm{n}})}{r_\mathbf{j,l}^3},\label{eq:Vdd}
\end{equation}
where $\hat{\mathrm{d}}^{(\mathbf{j})}$ is the dipole operator of atom $\mathbf{j}$. 
The projection of the dipole-dipole potential onto the Rydberg states manifold can be written as
\begin{equation}
H_\mathrm{int} = \frac{1}{2}\sum_{j\neq \mathbf{l}}PV_{\mathrm{dd}}^{(\mathbf{j,l})}P,
\end{equation} 
with the projection operator $P=\sum_{i_\mathbf{j},i_\mathbf{l}\in\lbrace 3,4,5,6,7\rbrace}|i_\mathbf{j}, i_\mathbf{l}\rangle\langle i_\mathbf{j},i_\mathbf{l}|$. Neglecting non-resonant processes (of the type $S_{1/2}S_{1/2}\to P_{1/2}P_{1/2}$), the Hamiltonian reduces to
\begin{equation}
H_\mathrm{int}=\sum_{\mathbf{j}\neq l}  \frac{C_3}{r_\mathbf{j,l}^3} c^\dagger_\mathbf{j} h_\mathrm{dd}(\theta_\mathbf{j,l},\phi_\mathbf{j,l}) c_\mathbf{l},
\end{equation}
with the shorthand notation $c_\mathbf{j}=[|5\rangle_\mathbf{j}\langle 3|,|6\rangle_\mathbf{j}\langle 3|,|5\rangle_\mathbf{j}\langle 4|,|6\rangle_\mathbf{j}\langle 4|]$ and the $4\times 4$ matrix $h_\mathrm{dd}(\theta_\mathbf{j,l},\phi_\mathbf{j,l})$ given by
\begin{equation}
h_\mathrm{dd}(\theta_\mathbf{j,l},\phi_\mathbf{j,l})=
 \left[ \begin{array}{cccc}
f_1(\theta_\mathbf{j,l}) &f_2(\theta_\mathbf{j,l},\phi_\mathbf{j,l})^* & f_2(\theta_\mathbf{j,l},\phi_\mathbf{j,l})  & -f_1(\theta_\mathbf{j,l}) \\
f_2(\theta_\mathbf{j,l},\phi_\mathbf{j,l}) &-f_1(\theta_\mathbf{j,l}) & -f_3(\theta_\mathbf{j,l},\phi_\mathbf{j,l})  & -f_2(\theta_\mathbf{j,l},\phi_\mathbf{j,l}) \\
f_2(\theta_\mathbf{j,l},\phi_\mathbf{j,l})^* &-f_3(\theta_\mathbf{j,l},\phi_\mathbf{j,l})^* & -f_1(\theta_\mathbf{j,l})  & -f_2(\theta_\mathbf{j,l},\phi_\mathbf{j,l})^* \\
-f_1(\theta_\mathbf{j,l})^* &-f_2(\theta_\mathbf{j,l},\phi_\mathbf{j,l})^* & -f_2(\theta_\mathbf{j,l},\phi_\mathbf{j,l})  & f_1(\theta_\mathbf{j,l}) 
\end{array} \right],
\end{equation}
with $f_1(\theta_\mathbf{j,l})=(1-3\cos^2\theta_\mathbf{j,l})/9$, $f_2(\theta_\mathbf{j,l},\phi_\mathbf{j,l})=e^{i\phi_\mathbf{j,l}}\sin 2\theta_\mathbf{j,l}/6$, $f_3(\theta_\mathbf{j,l},\phi_\mathbf{j,l})=e^{2i\phi_\mathbf{j,l}}\sin^2\theta_\mathbf{j,l}/3$.

Finally, an example for an excitation transfer process according to our dressing scheme is depicted in Figure~\ref{fig:app}(b).
\end{document}